\documentclass[aps,prl,twocolumn,groupedaddress,floatfix,superscriptaddress]{revtex4-1}
\usepackage{empheq}

\usepackage{graphicx}
\usepackage[hidelinks]{hyperref}
\usepackage{color}
\usepackage[T1]{fontenc}
\usepackage[utf8]{inputenc}
\usepackage{amssymb}
\usepackage{amsmath}
\usepackage{amsthm}
\usepackage{bbm}
\usepackage{mathtools}
\usepackage{float}
\usepackage{empheq}
\usepackage{rotating}

\def\tr{\operatorname{tr}}

\def\ket #1{\vert#1\rangle}

\def\tr{\mathop{\rm tr}\nolimits}

\def\lsection#1{\textit{#1}.---\hspace{-1em} }

\begin{document}

\title{Entanglement detection by violations of noisy uncertainty relations:\protect\\ A proof of principle}

\author{Yuan-Yuan Zhao}
\affiliation{
CAS Key Laboratory of Quantum Information, University of Science and Technology of China, Hefei 230026, P. R. China
}
\affiliation{School of Physics and Astronomy, Sun Yat-Sen University, Zhuhai, Guangdong, China 519082}
\affiliation{CAS Center for Excellence in Quantum Information and Quantum Physics,University of Science and Technology of China, Hefei, 230026, P. R. China}
\author{Guo-Yong Xiang}
\email[]{gyxiang@ustc.edu.cn}
\author{Xiao-Min Hu}
\author{Bi-Heng Liu}
\email[]{bhliu@ustc.edu.cn}
\author{Chuan-Feng Li}
\author{Guang-Can Guo}
\affiliation{
CAS Key Laboratory of Quantum Information, University of Science and Technology of China, Hefei 230026, P. R. China
}
\affiliation{CAS Center for Excellence in Quantum Information and Quantum Physics,University of Science and Technology of China, Hefei, 230026, P. R. China}
\author{René Schwonnek}
\email[]{rene.schwonnek@itp.uni-hannover.de}
\affiliation{Institut f\"ur Theoretische Physik, Leibniz Universit\"at Hannover, Germany}
\affiliation{Department of Electrical and Computer Engineering, National University of Singapore, Singapore}
\author{Ramona Wolf}
\email[]{ramona.wolf@itp.uni-hannover.de}
\affiliation{Institut f\"ur Theoretische Physik, Leibniz Universit\"at Hannover, Germany}

\begin{abstract}
It is well-known that the violation of a local uncertainty relation can be used as an indicator for the presence of entanglement. Unfortunately, the practical use of these non-linear witnesses has been limited to few special cases in the past. However, new methods for computing uncertainty bounds became available. Here we report on an experimental implementation of uncertainty-based entanglement witnesses, benchmarked in a regime dominated by strong local noise. We combine the new computational method with a local noise tomography in order to design noise-adapted entanglement witnesses. This proof-of-principle experiment shows that quantum noise can be successfully handled by a fully quantum model in order to enhance entanglement detection efficiencies.  
\end{abstract}

\pacs{123 456}

\maketitle

\section*{Introduction}
Entanglement is a crucial resource that enables quantum technologies like cryptography, computing, dense coding and many more. Hence, in addition to the challenging task of creating entanglement \cite{epr,aspect,zeilinger,zeilinger1,rfw89,satellite}, robust and practical verification schemes are a key requirement for unleashing the full power of these technologies.

\begin{figure}[t]
\includegraphics[width=0.95\linewidth]{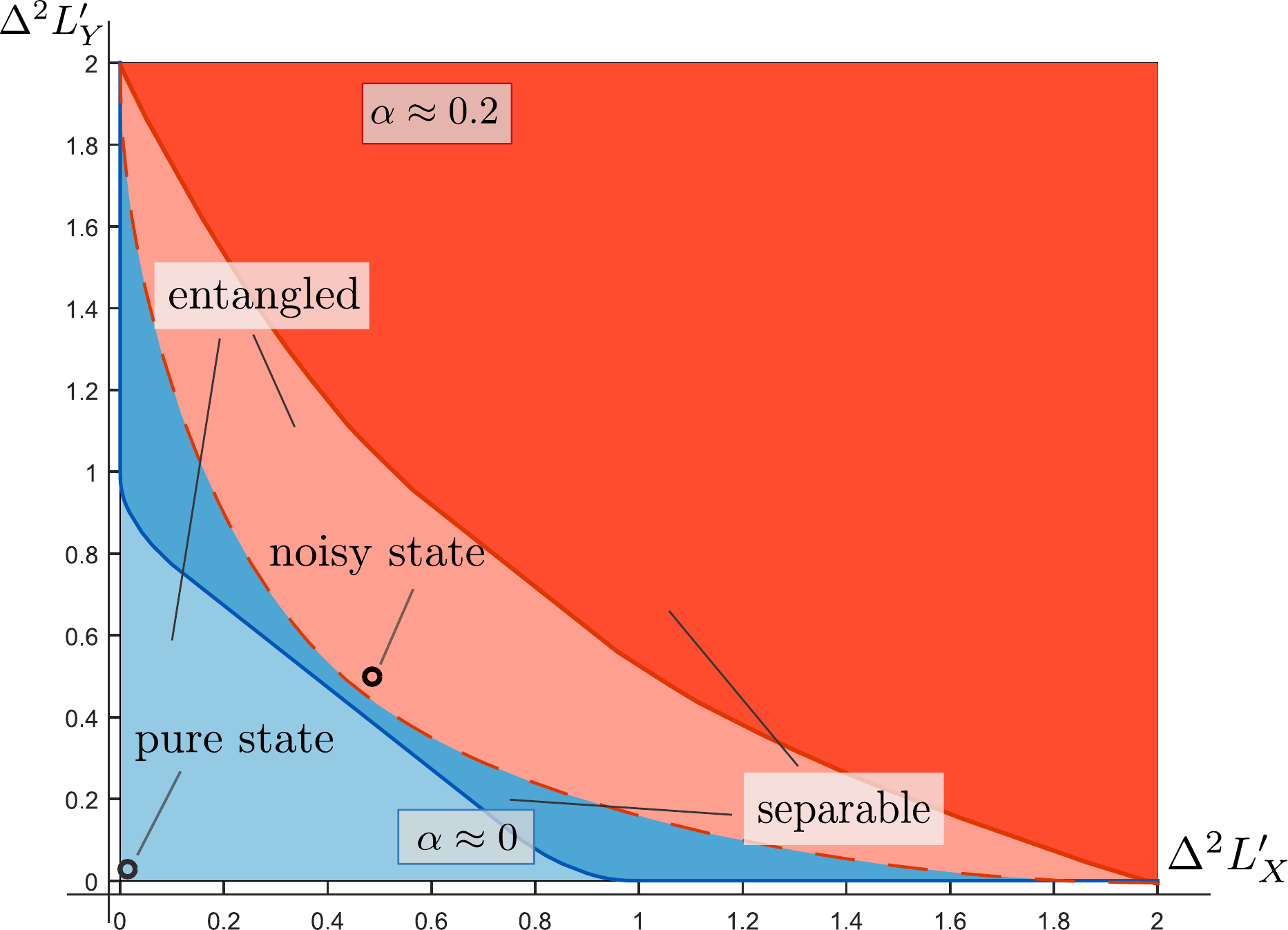}
\caption{\textbf{Uncertainty regions adapted to local noise.} We measure the variance of the total angular momentum of qutrit-qutrit states in two orthogonal directions. Since separable states have larger variances, the placement of an unknown quantum state in the above diagram allows us to conclude the presence of entanglement. The shape of these regions strongly changes by the influence of local noise sources. Depicted above for random spin-flip noise with estimated noise parameter $\alpha\approx 0$ (blue) and $\alpha\approx 0.2$ (red).
\label{fig:regions}}
\vskip-0.3cm
\end{figure}

In this letter we report a proof-of-principle experiment of noise-tolerant non-linear entanglement witnesses which are based on violations of local uncertainty relations. In our implementation, we detect photon encoded qutrit-qutrit entanglement by local measurements of orthogonal angular momentum components affected by strong noise originating from random spin-flips. We adapt our witnesses to this noise by an error estimation solely based on local measurements. We successfully benchmark our detection scheme in a noise regime where conventional witnesses fail to detect any entanglement at all (see Fig.~\ref{fig:regions}).

The existence of unavoidable uncertainties in any quantum
measurement process \cite{heisenberg,kennard,amu,jed,schrodingerucr,muff,ourentro,deutsch,blw,URu,schur,pati,alberto1,addi,sanders} is  one of the most characteristic implications of
quantum physics. It has been known for quite a while \cite{hofmann,guhne,guhne2,vardetect,ana,alberto} that any variance-based uncertainty relation 
\begin{align}
    \Delta^2 X +\Delta^2 Y \geq c
    \label{varugl}
\end{align}
on local measurements $X$ and $Y$ yields a non-linear entanglement witness. Unfortunately, in the past, this method could only be applied in a very limited context because explicit bounds in \eqref{varugl} were only known for few symmetrical cases (see e.g.~\cite{variances} for a list).
However, a method for computing bounds for \emph{general} $X$ and $Y$ recently became available \cite{variances,konrad}. 

For our experiment, we employ a modified version of this method in order to handle the influence of quantum noise entirely on a quantum level, whenever a quantum model of this noise is available.
In practice, we first motivate a noise model by theory and check it on a large range of tomographic test states afterwards. Thereby we concentrate on local noise sources \cite{streltsov}, since they are accessible in any LOCC setting. 

The wider scope of our proof-of-principle experiment are applications in long-range communication settings \cite{satellite,repeater1,repeater2}. Here, local noise typically turns out to be the actual limitation in practice, even though in theory, the exponential scaling of absorption is considered to be the limiting factor \cite{lutkenhaus,briegel}.

\section{Methods}
\lsection{Non-linear entanglement witnesses based on uncertainty relations}
In general, an entanglement witness is any separating functional $W(\rho)$ that can be used to distinguish an entangled state from separable states \cite{guhne}. More precisely, by taking the infimum of such a functional over all separable states, i.e.\ states of the form 
\begin{align}
    \rho = \sum_i p_i \rho^A_i \otimes \rho^B_i,\label{eq:sep}
\end{align}
we obtain a constant
	\begin{equation}
	\label{csep}
		c_{\text{SEP}}=\inf_{\rho\in\text{SEP}} W(\rho),
	\end{equation}
which allows us to witness entangled states, i.e.\ states that are not of the form \eqref{eq:sep} \cite{rfw89}: if we find that $W(\rho)< c_{\text{SEP}}$, we immediately know that $\rho$ cannot be a separable state and is therefore entangled. 

Typically linear functionals $W$ are considered. In contrast, the entanglement witnesses we are using in our experiment are based on variances of measurements $X$ and $Y$ and therefore non-linear. Explicitly, we use weighted uncertainty sums given by the functional
	\begin{equation}
		V(X,Y,\rho)=\lambda\Delta_\rho^2 X+\mu\Delta_\rho^2 Y,
		\label{eq:witness}
	\end{equation}
where $\Delta_\rho^2 X=\langle X^{(2)}\rangle-\langle X^{(1)}\rangle^2$. Here, $X^{(N)}$ are the so-called moment operators given for general POVMs \cite{teiko} as follows: Let $x$ be an outcome of $X$ and $P_X(x)$ the corresponding POVM element, then the $N$-th moment operator of $X$ is given by $X^{(N)}=\sum_x x^N P_X(x)$. 

 From an experimental perspective, our witnesses have the advantage that only the first two moments of an outcome distribution are needed.  In contrast to the full outcome statistics, these moments can be estimated with much higher precision given a finite sample.

From a theoretical perspective, the corresponding constant $c_{\text{SEP}}$ plays the role of a sum of local uncertainty bounds:
Consider a setting where two parties, named Alice and Bob, can both choose between two measurements $X$ and $Y$. We then have four local measurements, $X_A$ and $Y_A$ for Alice and $X_B$ and $Y_B$ for Bob. Globally, the $N$-th moment of the $X$-measurement is then given by
	\begin{equation}
		X^{(N)}=\sum \left(x_A+x_B\right)^N P_{X_A}(x_A)\otimes P_{X_B}(x_B)
	\end{equation}
and similar for $Y^{(N)}$.

The chosen witness has some convenient properties with regard to calculating the infimum in equation~(\ref{csep}): Since it is a concave function which we want to minimize over a convex set (the set of all separable states), the optima are obtained at extreme points. However, the extreme points of the set of separable states are pure states, i.e.\ we only have to find the infimum over all product states. The variance is then additive:
	\begin{equation}
		\Delta_{\rho_A\otimes\rho_B}^2 X=\Delta_{\rho_A}^2 X_A+\Delta_{\rho_B}^2 X_B,
	\end{equation}
which leads to the following expression for the uncertainty bound:
	\begin{align}
		c_{\text{SEP}}&=\inf_{\rho_A\otimes \rho_B} V(X,Y,\rho)\\
		              &=\inf_{\rho_A} V(X_A,Y_A,\rho_A)+\inf_{\rho_B}V(X_B,Y_B,\rho_B).
	\end{align}
Hence, we only need bounds on the functional $V$ with respect to local measurements and states. \\[0.1cm]

\begin{figure}[h]\centering
\includegraphics[width=0.6\linewidth]{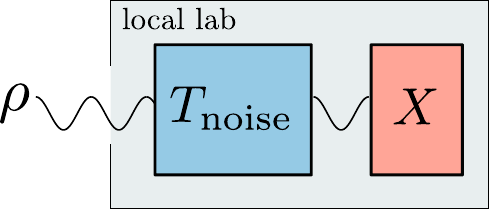}
\caption{\label{local_noise} \textbf{Schematic view on local noise.} Before a particle is measured, it is disturbed by noise that is modelled by a channel $T_{\text{noise}}$. Since $T_{\text{noise}}$ acts locally, its characteristics can be estimated in advance in order to construct error adapted entanglement witnesses.}
\end{figure}

\lsection{Local noise}
For the design of noise robust entanglement witnesses, we abstract the action of any local noise source as depicted in Fig. \ref{local_noise}: Whenever a particle enters a local lab, it is firstly affected by local noise, i.e.\ it has to cross a channel $T_{\text{noise}}$, before it hits an idealized detector $X$. 

From the perspective of the Schrödinger picture this results in a disturbed state $\rho'=T_{\text{noise}}[\rho]$, which then results in disturbed measurement outcomes with 
disturbed moments:
\begin{align}
\tr\left(T_{\text{noise}}[\rho] X^{(N)}\right).
\end{align}
Generically, these moments lead to an increase of the uncertainty $\Delta^2_{\rho'}X$, which reflects the negative effect of the noise. Hence entanglement, which was present in the initial state $\rho$, may no longer be detectable by a witness based on the local uncertainty of the ideal measurement $X$.

At this point we should keep in mind that, from a mathematical perspective, the disturbed state $\rho'$ still contains a lot of information about the undisturbed input $\rho$. Here one strategy could be to implement an error-correcting quantum channel $T_{\text{noise}}^{-1}$, which is unfortunately not practical: beside the fact that this would demand a very high level of quantum control, a full recovery of an \textit{unknown} $\rho$ is usually not possible since the inverse of a noise channel is typically not a $CP$-map which means that it is not a valid quantum operation.


This fundamental shortcoming can be circumvented by representing the noise in the Heisenberg picture: From this perspective, only the local detectors are affected by the noise. Here, we can assume that the characteristics of an ideal detector $X$ are \textit{well known} in advance, such that we can directly describe noisy measurements $X'$ by a POVM with elements 
\begin{align}
P_{X'}=T^*_{\text{noise}}[P_X(x)],
\end{align}
and moments 
\begin{align}
\tr\left(\rho T^*_{\text{noise}}[X^{(N)}]\right)=\tr\left(\rho \sum_{x} x^N T^*_{\text{noise}}[P_X(x)]\right).
\end{align}
Here, the local noise is compensated on a classical level when we adapt our entanglement witnesses by using the correct local uncertainty bounds for disturbed measurements $X'$. In practice this demands us to (i) collect information on the local noise in order to come up with a valid error model and (ii) compute the corresponding uncertainty bounds. We achieved the first task by performing measurements on (local) test-states and the second by using the recently developed algorithm from \cite{variances}, which can also be applied to arbitrary POVMs.\\[0.1cm]

\section{Experiment}
\begin{figure}[htp]
    \centering
    \includegraphics[width=1\linewidth]{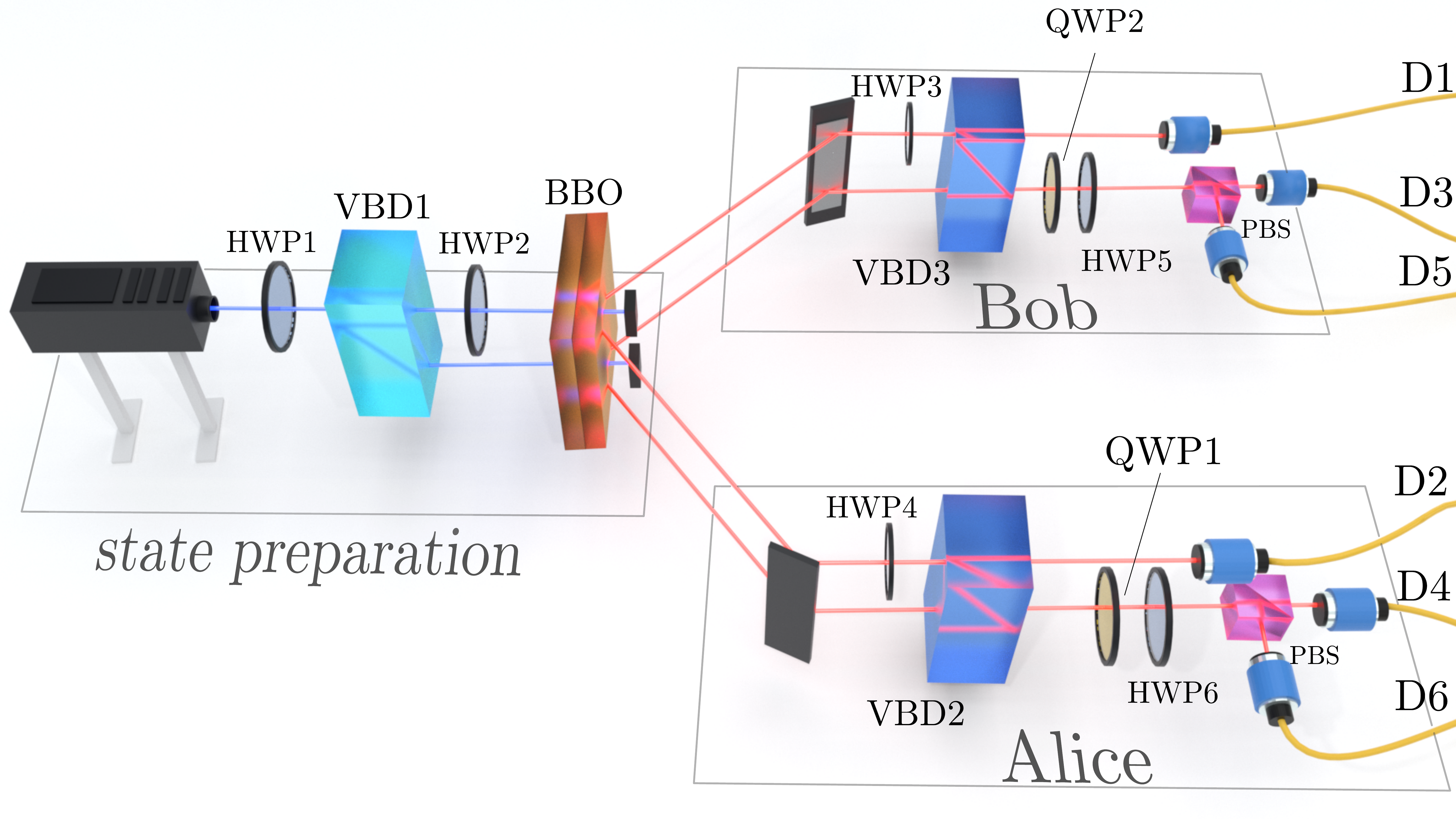}
    \caption{{\bf Experimental setup.} The pump state is prepared as $\psi_p=\frac{1}{\sqrt{3}}(\ket{H}_u+\ket{V}_u-\ket{V}_l$ using the beam displacer (VBD1) operated at $404$ nm and the HWPs (HWP1 and HWP2), where H (V) stands for the horizontal (vertical) polarization and the u (l) stands for the upper (lower) path. Then the pump photon splits to the entangled photon pairs via the spontaneous parametric
down conversion process. So we get the three dimensional singlet state in Eq.~\eqref{eq:singlet}. Then, Alice and Bob construct the measurement apparatus using the BD@$808$nm and the wave plates oriented at the specific angles. The details can be found in the supplementary materials.}
    \label{fig:setting}
\end{figure}

\lsection{Setup}
In our implementation, entanglement is realized by so called path polarization hybrid states \cite{qutritEx1,qutritEx2}. Due to many developments in recent years, the path and the polarization degrees of freedom of a photon can be controlled easily and efficiently. 

We use the setup as sketched in Fig.~\ref{fig:setting} to prepare the singlet state
\begin{align}
   \psi_{\text{singl}}=\frac{1}{\sqrt{3}}\big(\ket{02}+\ket{20}-\ket{11}\big).
   \label{eq:singlet}
\end{align}
We detect the entanglement of this state based on measurements of spin-1 angular momentum components $L_X$ and $L_Y$.\\[0.1cm]

\lsection{The noise model}
In our experiment we probe qutrit-qutrit entanglement detected by local measurements of the spin-1 components $L_X$ and $L_Y$. 
In order to benchmark the performance of our method in a regime where the conventional criteria fail to work, we actively add local noise to our measurements.

We generate this noise by applying a random sequence of local spin-flips within the $L_X-L_Y$ plane.
For a noise parameter $\alpha$ which corresponds to an effective spin-flip probability this implements a channel 
\begin{align}
T^*_{\text{noise}}[P]:= \frac 12 \left( (2-\alpha) P + \alpha U^*_{\text{flip}} P U^{\phantom{*}}_{\text{flip}} \right), \label{tnoise}
\end{align}
where an individual spin-flip is described by the unitary  
$U_{\text{flip}}:= e^{-i\pi L_Z}$.

\begin{figure}[ht]
\includegraphics[width=0.95\linewidth]{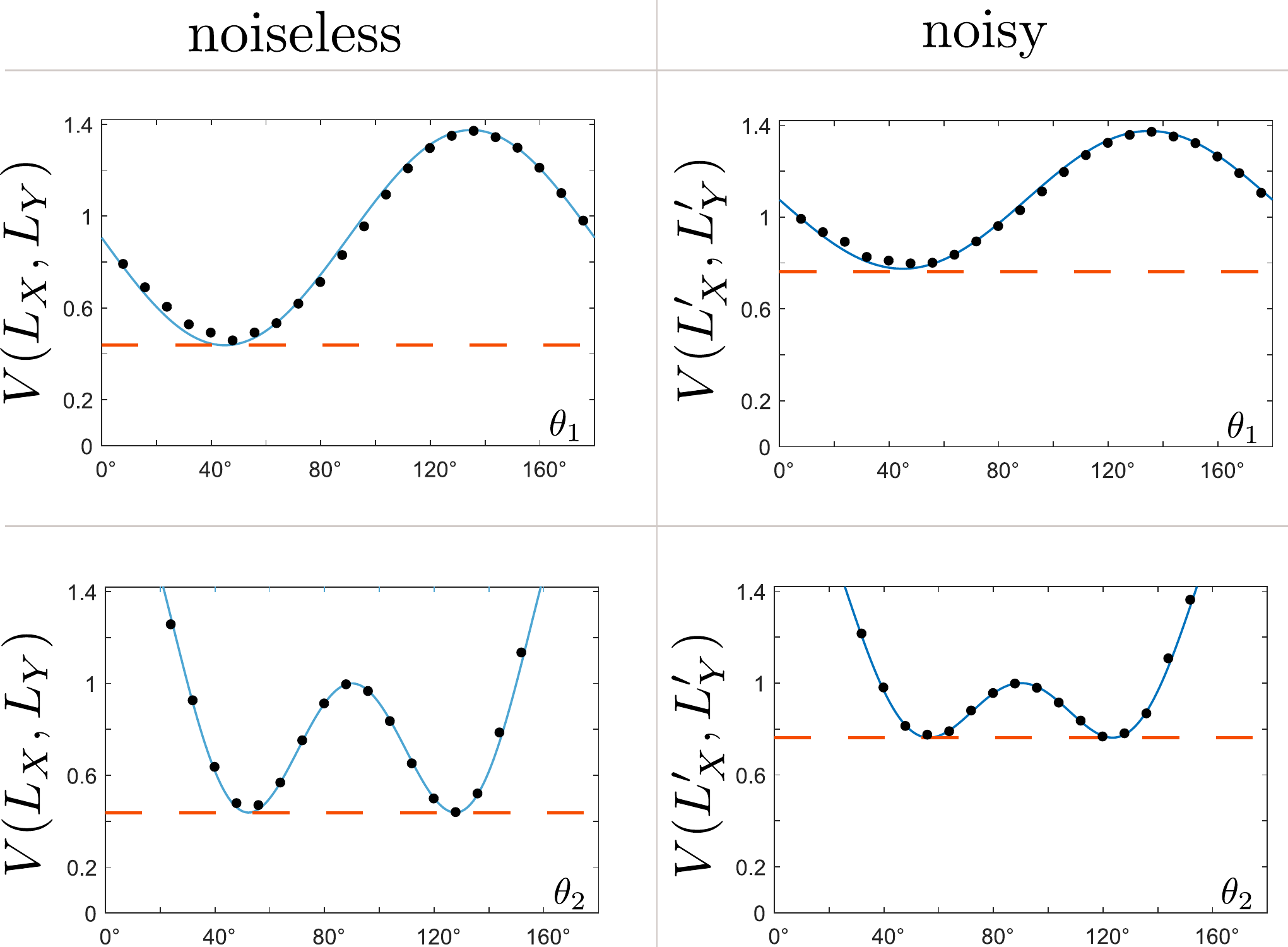}
\caption{\textbf{Calibration of the measurement box.}
Local uncertainty sums for noiseless ($V(L_X,L_Y)$) and  noisy ($V(L_X',L_Y')$) observables measured on two sequences of test states $\psi_{\text{test}}(\theta_1,\theta_2)$.
Upper row: prepared in $2\pi /45$ steps for $0 < \theta_1< 180^{\circ}$ and $\theta_2=23.3^{\circ}$. Lower row: prepared
in $2\pi /45$  steps from $0 < \theta_2< 180^{\circ}$ and $\theta_1=28^{\circ}$. \\
For each state we collect about $20,000$ photons in total. The statistical errors caused by the fluctuation of the coincidence count stay below $0.004$ and $ 0.01$, which is too small to be depicted in the figures.\label{noisetest} 
}
\end{figure}

For this channel the operators $L_X'^{(1)}$ and $L_Y'^{(1)}$, corresponding to the first moments of our noisy detectors, 
are given by 
\begin{align}
L_X'^{(1)}&=\frac{1}{\sqrt{2}}
\begin{pmatrix}
0&1-\alpha&0\\
1-\alpha&0&1-\alpha\\
0&1-\alpha&0
\end{pmatrix} ,
\\
L_Y'^{(1)}&=
\frac{i}{\sqrt{2}}
\begin{pmatrix}
0&\alpha-1&0\\
1-\alpha&0&\alpha-1\\
0&1-\alpha&0
\end{pmatrix},
\end{align}
whereas the operators for the second moments stay unchanged, i.e.\ we have 
\begin{align}
L_X'^{(2)}=\frac{1}{2}
\begin{pmatrix}
1&0&1\\
0&2&0\\
1&0&1
\end{pmatrix} ,
\quad
L_Y'^{(2)}=\frac{1}{2}
\begin{pmatrix}
1&0&-1\\
0&2&0\\
-1&0&1
\end{pmatrix} .
\end{align}
Given this noise model we probe the actual local noise by checking the predicted local uncertainty relations on a set of test states
\begin{align}
    \psi_{\footnotesize{\rm test}}= \sin\theta_1 \cos\theta_2 \ket{0} + \cos\theta_1 \ket{1} +
\sin\theta_1 \sin\theta_2\ket{2}.
\end{align}
The results of this are depicted in Fig.~\ref{noisetest}.

In principle, our method provides the ability to incorporate more sophisticated noise models than the presented one. However, for this experiment, it turns out that the dominant part of the noise regime can be well explained (see Fig.~\ref{noisetest}) by the action of the channel \eqref{tnoise}, with a spin-flip noise parameter $\alpha\approx 0.2$, estimated from experimental data. \\[0.1cm]

\lsection{Entanglement detection}
\begin{figure}[ht]
    \centering
    \includegraphics[width=0.85\linewidth]{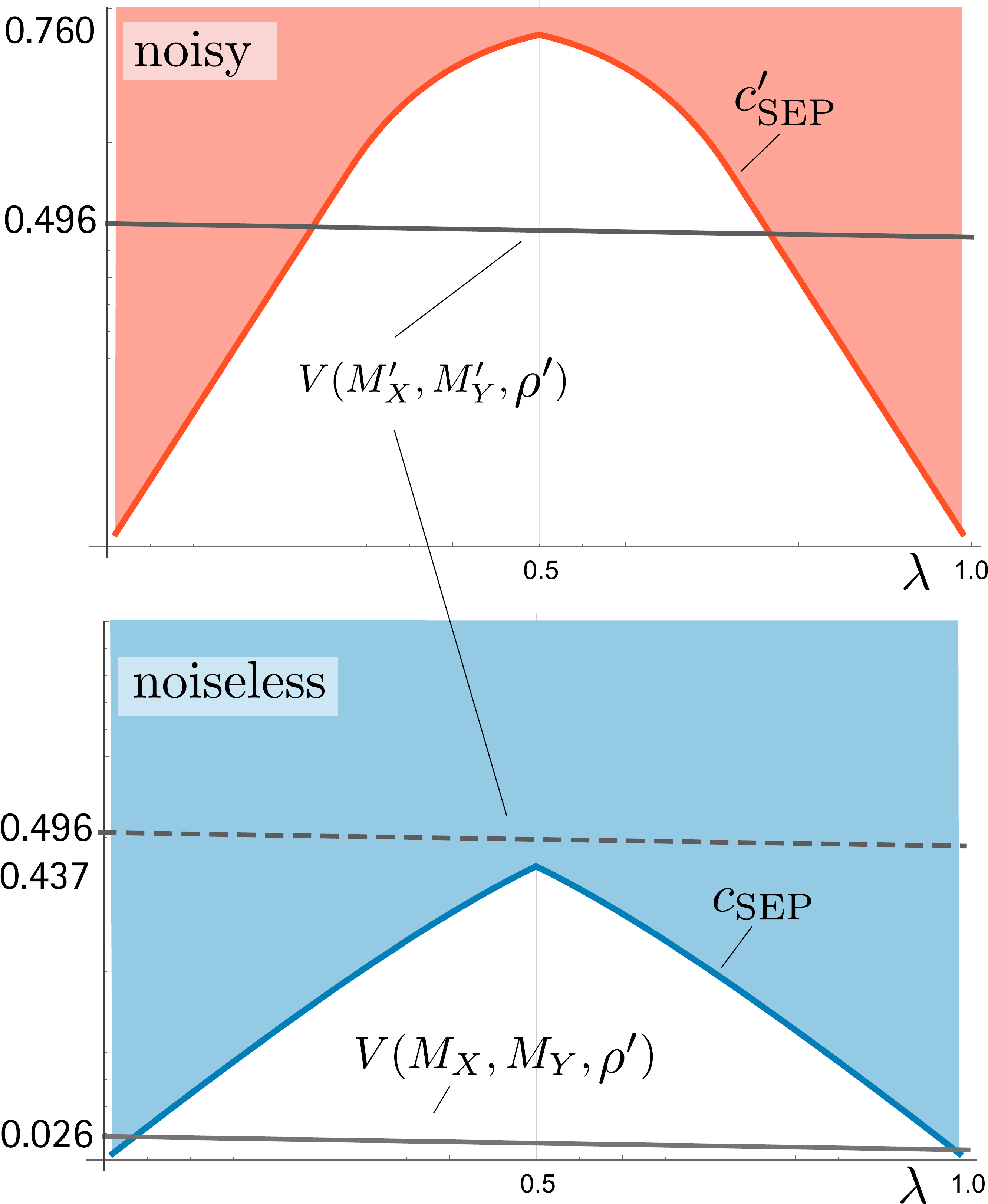}
    \caption{\textbf{Entanglement detection.} Uncertainty bounds $c_{\rm SEP}$ (blue line) for the noiseless and $c'_{\rm SEP}$ (red line) for noisy measurement and the variance functional $V$ (black solid and dashed lines) evaluated on the measured state, parameterized by $\lambda$. Whenever a point on the black line lies below the uncertainty bound, i.e.\ outside of the colored region, entanglement is witnessed. }
    \label{fig:results}
\end{figure}
Via the procedure described above, we can do measurements on a state $\rho'$ which is, in principle, unknown but close to the singlet state~\eqref{eq:singlet}.  The singlet state has the highest entanglement which yields the highest violation of the computed bounds, hence it is favourable to use this state for the purpose of benchmarking. 

The operators we measure are
	\begin{align}
	    M_i^{(1)}&=L_i^{(1)}\otimes\mathbb{I}+\mathbb{I}\otimes L_i^{(1)}, \\
		M_i^{(2)}&=L_i^{(2)}\otimes\mathbb{I}+2 L_i^{(1)}\otimes L_i^{(1)} + \mathbb{I}\otimes L_i^{(2)},
	\end{align}
for $i=X,Y$, i.e.\ the first two moments of the total angular momentum (see also \cite{spinsqueezing} for similar applications on BECs). The noisy operators ${M'_i}^{(N)}$ are defined analogously.

The measurement is performed once without noise and once with noise. The major advantage of our scheme is that we can already test a variety of witnesses, i.e.\ all values of $\lambda$ and $\mu$ in our witness functional~\eqref{eq:witness}, 
by performing a measurement of only two spin components ($M_X$ and $M_Y$).
More precisely, we measure a variance tuple ($\Delta_{\rho'}^2 M_X,\Delta_{\rho'}^2 M_Y$), from which the values of the variance functional $V(M_X,M_Y,\rho')$ can be computed for all $(\lambda,\mu)$. Whenever this functional violates an uncertainty bound, that is we observe 
 \begin{align}
    c_{\rm SEP} \geq V(M_X,M_Y,\rho')
 \end{align}
for some choice of parameters, entanglement is detected.

The results of this are shown in Fig.~\ref{fig:results}, where we used, without loss of generality, the parameterization  $\mu=1-\lambda$. 
We detect entanglement with witnesses in the ranges $\lambda \in \{0.028,0.985\}$ and  $\lambda \in \{0.250,0.755\}$, for the noiseless and the noisy case, respectively. Remarkably, we observe for the noisy case that, even for the highly entangled state we used here, the non-adapted witness would fail to detect any entanglement (for any $\lambda$), as shown by the placement of the dashed line in Fig.~\ref{fig:results}.

Furthermore, the strongest witness for this particular state is given by the parameters $\lambda=\mu$. Here we obtain the uncertainty bounds
\begin{align}
   \frac{1}{2} \Delta^2 M_X + \frac{1}{2} \Delta^2 M_Y \geq \frac{7}{16} \approx 0.4375 = c_{SEP} \label{witoutnoise}
\end{align}
and 
\begin{align}
   \frac{1}{2} \Delta^2 M_X' + \frac{1}{2}  \Delta^2 M_Y' \geq 0.7614 = c_{SEP}' \label{witnoise}
\end{align}

\noindent
whereas the corresponding witness functionals only attain the values
\begin{align}
    V(M_X,M_Y,\rho')=0.021\pm0.003\leq c_{\rm SEP}
\end{align}
and 
\begin{align}
    c_{\rm SEP}\leq V(M_X',M_Y',\rho')=0.492\pm0.018 \leq c'_{\rm SEP}, \label{valnoise}
\end{align}
which again shows that, given the observed data \eqref{valnoise}, the witness \eqref{witnoise} detects entanglement whereas the witness \eqref{witoutnoise} fails.\\[0.1cm]

\section{conclusion}

We described a proof of principle experiment that demonstrates that high precision experimental techniques and improved computational methods can be merged to turn a simple idea \cite{hofmann} into a practical technology. 

The local noise in this experiment has a relatively simple appearance, but our techniques are not limited to this:  A clear direction for future work is to use this scheme for an improvement of existing entanglement distribution setups, especially in long range settings. 
It is also worth mentioning that our scheme can be extended to settings with more than two local measurements as well. Beside experimental challenges, this also demands new numerical methods. For this case, the corresponding  uncertainty relations are way less investigated than in the case of two measurements. Here, new numerical methods and experimental tests would contribute a lot to the understanding of this case.

\quad
\\
\begin{acknowledgments}
  
R.S. and R.W. thank Reinhard F. Werner, Tobias J. Osborne, and Deniz E. Stiegemann for fruitful discussions and critically reading our manuscript. Y.Y.Z. thanks Yu Guo's help in the experiment.
R.S. and R.W. also acknowledge the financial support given by the RTG 1991 and the CRC 1227 DQ-mat funded by the DFG, the collaborative research project Q.com-Q funded by the BMBF and the
Asian Office of Aerospace RD grant FA2386-18-1-4033. 
The work at USTC is supported by the National Natural Science Foundation of China under Grants (Nos.11574291, 11774334, 61327901,11874345 and 11774335), the China Postdoctoral Science Foundation (Grant No. BH2030000036), the National Key Research and Development Program of China (No.2017YFA0304100), and the Key Research Program of Frontier Sciences, CAS (No.QYZDY-SSW-SLH003), Anhui Initiative in Quantum Information Technologies.

\end{acknowledgments}
\bibliography{library}
\end{document}